\documentclass[12pt]{iopart}

\usepackage[english]{babel}
\usepackage[T1]{fontenc}

\usepackage{lmodern}

\usepackage{iopams}
\usepackage{graphicx}
\usepackage{pstricks}
\usepackage{pst-node}
\usepackage{amsmath}
\usepackage[normalem]{ulem}

\usepackage{cite}      
\usepackage{graphicx}

\usepackage{amsmath}   
\usepackage{url}
\usepackage{subfigure}
\usepackage{multirow}
\usepackage{pstricks}
\usepackage{pst-node}
\usepackage{multirow}
\usepackage{rotating}
\usepackage{amsfonts}
\usepackage{amssymb}
\usepackage{exscale}
\usepackage{upgreek}
\usepackage{textcomp}
\usepackage{epsfig}
\usepackage[T1]{fontenc}
\usepackage{perpage}

\begin{document}

\newcommand{\TM}{{\mathrm{T}\Omega}}
\newcommand{\TxM}{{\mathrm{T}_x\Omega}}
\newcommand{\TstarM}{{\mathrm{T}^{*}\Omega}}
\newcommand{\TstarxM}{{\mathrm{T}^{*}_x\Omega}}

\newtheorem{mydef}{Definition}
\newtheorem{mytheorem}{Theorem}
\newtheorem{myremark}{Remark}
\newtheorem{myrp}{RP}
\newtheorem{myconjecture}{Conjecture}

\title[]{Semantics of HTS AC loss modelling: Theories, models and experiments}

\author{V Lahtinen and A Stenvall}

\address{Electrical Engineering, Tampere University, PO Box 1001, 33014 Tampere University, Finland}

\ead{valtteri.lahtinen@tuni.fi}

\begin{abstract}

Computer-assisted modelling is an essential approach to design new devices. It speeds up the process from the initial idea to an actual device and saves resources by reducing the number of built prototypes. This is also a significant practical motivator behind scientific research in contemporary high-temperature superconductor (HTS) AC loss modelling. However, in the scientific literature in this field, consistent practices about modelling terminology have not been established. Then, it is up to the reader to decide, what is the true intent and meaning of the authors. Consequently, the interpretation of such literature might be very much reader-dependent. Moreover, an inseparable part of the whole modelling process is the development of modelling approaches and numerical methods and comparing the predictions obtained via modelling to experimentally achieved results: It is commonplace to discuss the accuracy of modelling results or the validation of a model. In this paper, we discuss the terminology related to theories, models and experiments in the context of HTS AC loss modelling. We discuss the recursive nature of theories and models in this context, discuss the compatibility of discrete formulations of physics utilized in our field with the corresponding continuum descriptions, as well as with intuition, and interpret the perceived meaning of validation of a self-consistent model, shedding light on the relationships between theories, models and measurements. We present our view on understanding these relations in the familiar context of AC losses in HTS. As a result, we end this paper with four conjectures describing our views.

\end{abstract}


\maketitle

\section{Introduction}

Computer simulations to predict AC losses in high-temperature superconductors (HTS) are required in the design phase of devices made out of such materials. This need has given rise to a niche subfield of science dedicated to numerical modelling of AC losses \cite{HTSWorkgroup}. However, perhaps due to the relatively young age of this field, consistent practices about related terminology have not been established, which has the consequence of disconcerting author-dependency and reader-dependency of the scientific literature. The need for consistency in the terminology, as well as the need for the recognition of some of the more philosophical aspects of the science conducted in this field were recently addressed in the \emph{Sixth International Workshop on Numerical Modelling of HTS} in Caparica, Portugal in the summer of 2018 \cite{HTSWorkshop}. This paper is a natural outgrowth of the discussions initiated there.

\begin{quote}
\emph{``To understand science is to know how scientific models are constructed and validated.''} (David Hestenes \cite{Hestenes})
\end{quote}

\subsection{Motivation and background}

As the criteria of successfulness of researchers are these days largely of quantitative nature, a natural consequence is that the perceived value of an incremental contribution has come closer to that of an average or even seminal contribution. This is also the case in our field. The multitude of papers published makes it arguably more and more difficult to pick out the important contributions among the published articles -- whether the paper is application-oriented, presents new modelling methodology or discusses a completely new approach to a problem. This may lead to inefficient use of available overall resources. To deal with this in the field of HTS AC loss modelling, we believe organization and synthesis is useful. With this paper, we aim to start this work: This paper seeks to provide synthesis by unifying the terminology and considering some of the fundamental issues related to such scientific research. In this sense, this is a sister paper for \cite{StenvallLahtinenASC2018} presented in the Applied Superconductivity Conference in Seattle, Washington in the fall of 2018 \cite{ASC}. Also, this paper complements a recent effort with somewhat similar goals \cite{GrilliSirois}.

In our field of study and more generally, in engineering and natural sciences, the distinction between a theory and a model is often blurred in the everyday practice of scientific discussion, and the terms are used somewhat interchangeably. The nature and difference of these two related notions is also a central research issue in the philosophy of science. Modern literature encases a multitude of different views in terms of e.g. the related ontology and semantics: A comprehensive review on such viewpoints can be found in \cite{Frigg}. However, our view is motivated by pragmatic benefits and provides us with a mental framework for the process of forming mathematical models of physical theories, and thus, yields also a clear distinction between the two concepts, suitable for our field. However, we also suggest that the distinction between a theory and a model is not an absolute notion but relative with respect to the chosen abstraction level. We argue that acknowledging this recursive chain structure of theories and models benefits the scientific research in our field.

The mathematical models we utilize when modelling HTS AC losses are typically formalized in continuum either utilizing point-wise quantities and (partial) differential equations describing the phenomena or macroscopic quantities and algebraic equations. This yields the need for a \emph{discretization}, a way to finite-dimensionalize the space from where to search for the solution. Using different discretization methods and formulating the utilized models in different ways, one obtains discrete descriptions that preserve different aspects of the original continuum-based description. The \emph{compatibility} of a discretization in this sense should be understood as a key issue also in HTS AC loss modelling. 

Arriving at a discrete description of the phenomenon we are modelling, suitable for a computer and perhaps compatible with the corresponding continuum description in some sense, we can make predictions. The question arises: How do we know if the predictions we make are any good? We must compare the predictions of our models to predictions of other models, and perhaps more preferably, to experimental data. This brings us to the notion of \emph{model validation}. The central question here is: What is the relation between simulation results and experiments, and in what sense, if any, does such \emph{model validation} validate a model? This is also one of the central issues of this paper.

\subsection{The structure of the paper}

In section~2, we discuss the recursive nature of theories and models. Then in section~3, we consider how we formulate our models and discuss also the \emph{compatibility} of our discrete descriptions with the corresponding continuum descriptions. In particular, we will discuss an algebraic formulation not so well-established in the AC loss modelling community in more detail. In section~4, we take on the issue of relating modelling to experiments, and finally in section~5 we conclude with four conjectures, reflecting the perspective on modelling we advocate throughout the paper.

\section{Theories vs. models: a recursive system}

In this section, we discuss the nature of and relationship between \emph{theories} and \emph{models}. In mathematics, a theory is a collection of axioms, taken as a starting
point for deduction, that concern a set of undefined elementary entities. Theories of physics build upon the frameworks provided by mathematical theories by postulating some defining properties for the mathematical objects of a mathematical theory. Naturally, such a theory is typically based on physical intuition and mathematics is the way to formalize this intuition. As our field of interest is AC loss modelling of HTS, we shall restrict our discussion to \emph{mathematical theories and models of physics}. 

\subsection{Maxwell's theory}

Eventuallly, modelling physical phenomena utilizing models about nature is about solving equations. However, we know that models and theories utilized are something more than the formulas used to represent them. In the end, we want to represent something from the real world utilizing mathematics and often computers, to predict natural phenomena. Still, models seem to arise from axioms independently of how nature behaves. And even more clearly, things from the real world do not obey any models. As an example, real superconductors do not obey the Bean's model \cite{Bean}, or the power-law model \cite{Bruzzone}, which describe their transition between superconducting and normal conducting states, but indeed, both models are very useful in predicting the behaviour of superconductors in certain situations.

But how do we form models from theories? And what are (mathematical) theories of physics in the first place? To take an example familiar to an AC loss modeller, consider Maxwell's theory of electromagnetism, which, in our field, is typically stated as

\begin{equation}\label{Faraday}
\nabla \times {\mathbf E} = -\partial_t {\mathbf B},
\end{equation}
\begin{equation}\label{Ampere-Maxwell}
\nabla \times {\mathbf H} = {\mathbf J} + \partial_t {\mathbf D},
\end{equation}
\begin{equation}\label{Gauss1}
\nabla \cdot {\mathbf B} = 0,
\end{equation}
\begin{equation}\label{Gauss2}
\nabla \cdot {\mathbf D} = q,
\end{equation}
with the constitutive laws
\begin{equation}\label{constitutives}
{\mathbf D} = \epsilon {\mathbf E}, \quad {\mathbf B} = \mu {\mathbf H}, \quad {\mathbf E} = \rho {\mathbf J},
\end{equation}
In these equations, ${\mathbf E}$, ${\mathbf D}$, ${\mathbf H}$ and ${\mathbf B}$ are the electric field intensity, electric flux density, magnetic field intensity and magnetic flux density, respectively. These are mathematical beings called \emph{vector fields}. The electric charge density $q$ is a scalar field. Moreover, these fields are connected through the material parameters $\epsilon$, $\mu$ and $\rho$. Typically, in our field, we work within the magnetoquasistatic regime, where we assume $\partial_t {\bf D} = 0$. 

The standard approach for 3-D modelling is to assume that the modelling domain in which we are solving \eqref{Faraday}-\eqref{constitutives} is embedded in $\mathbb{R}^3$. Furthermore, we typically assume that the \emph{metric tensor} of our modelling domain is Euclidean, which in $\mathbb{R}^3$ induces the standard norm that can be represented as $\|\mathbf x\| = \sqrt{\sum^{3}_{i}{x^{2}_{i}}}$, where $x_i$ are the components of the vector ${\mathbf x}$. In fact, if we read the equations \eqref{Faraday}-\eqref{constitutives} literally, we have even fixed the coordinate system to be Cartesian: If e.g. $\nabla \times {\bf H}$ means literally the cross product of the formal vector of partial derivatives and the vector field ${\bf H}$, Maxwell's equations as presented above do not hold for e.g. spherical coordinates. However, if $\nabla \times {\bf H}$ is merely a notational tool to express the $\mathrm{curl}$ of ${\bf H}$, then we are not necessarily bound to Cartesian frames. Nonetheless, Maxwell's equations in the form presented above require the structures of metric tensor and \emph{volume form}, as the definitions of curl and divergence require them, respectively \cite{Frankel}. 

\subsection{Theory-model recursion -- Is Maxwell's theory a theory or a model?}

It is commonly agreed that Maxwell's theory of electromagnetism as presented above is a theory. To obtain models from this theory, we can instantiate the free pieces in it: For example, we are left to decide the material properties, the modelling domain $R \subset {\mathbb R}^3$ and source fields of this theory as we wish to instantiate a model, and use it for modelling the electromagnetic phenomena in a given real-world situation. This is how we make models from theories: we instantiate the free pieces, the \emph{structures} left uninstantiated in the theory. However, we could have represented the same equations in a more general setting on a Riemannian manifold $\Omega$ utilizing differential geometry \cite{Hehl}. From this perspective, the theory presented above is merely an instance of this more general theory -- its model. We would have not had to instantiate the modelling domain to be Euclidean: such a domain is just an instance of a Riemannian manifold.\footnote{The formalism of differential geometry does not itself take a stand on which abstraction level the theory we are dealing with lives. However, it makes the different abstraction levels more transparent, allowing to e.g. differentiate between the metric-dependent and metric-independent parts of the theory or model. Maxwell's equations are inherently metric-independent and the constitutive equations are inherently metric-dependent. For a brief introduction to AC loss modelling in the framework of manifolds and differential geometry, see \cite{LahtinenSUST}.}

The above example suggests, that the difference between a theory and a model is not an absolute notion. Instead, whether to call such a set of defining properties a theory or a model is a decision left for the modeller. Hence, we can think of Maxwell's theory of electromagnetism, as presented above, either as a theory or a model of a more general theory. Note how, when understood in this sense, e.g. the power-law model and the critical state model are indeed different models of the same theory, (magnetoquasistatic) Maxwell's theory. They differ in their instantiation of the constitutive equation connecting ${\bf E}$ and ${\bf J}$.  

The key difference between a theory and its model is that the model instantiates something in the theory. In the science of modelling, it can sometimes be useful to go towards more abstract to look for something new or to a ``silo'' where no one has ever been yet. It is of crucial importance to acknowledge the abstraction level one is working at: what there is to instantiate and what can be found by raising the abstraction level. As an example, consider again Maxwell's theory of electromagnetism. The whole theory is implied by the theory of quantum electrodynamics but not vice versa \cite{Feynman}. The whole classical theory of electromagnetism is thus reducible to a mere consequence, or a special case, of a more fundamental theory. In a related manner, by raising the abstraction level, we can \emph{unify} several seemingly different theories or models under one, more abstract entity containing them.\footnote{As examples from different fields of science, the wide applicability of monoidal categories in, e.g., computation, physics and engineering is an archetypical manifestation of this \cite{RosettaStone}, \cite{Coecke}, \cite{Lahtinen}. Or, looking at Maxwell's theory again, Maxwell's equations are merely an instance of the more general Yang-Mills equations (see e.g. \cite{BaezMuniain}) -- one model of this theory.}.

\subsection{Concluding remarks}

\begin{figure}[!t]
\center
\includegraphics[scale=0.75]{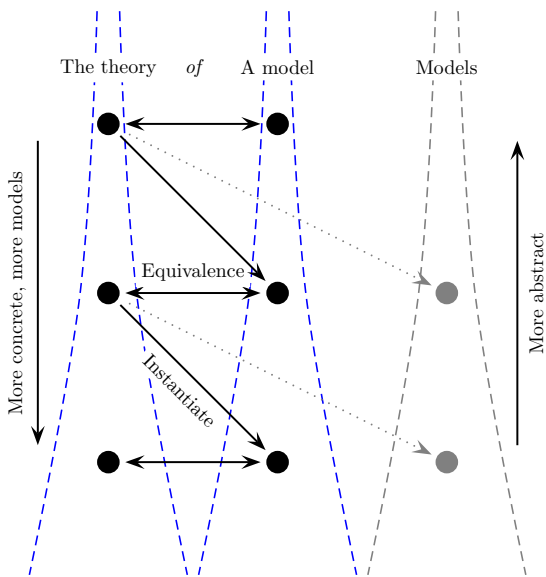}\caption{Theory-model recursion. A model is formed by instantiating structures left uninstantiated in a theory. This model can still leave structures uninstantiated and looking from this level of abstraction, it is a theory for a model at a lower abstraction level.}
\label{fig:theory-model}
\end{figure}

We propose that theories and models form a recursive system, as shown in \fref{fig:theory-model}. Instantiating the free structures in a theory, one travels towards a more concrete description and forms models of the theory. On the other hand, free pieces may still exist at this abstraction level and in this sense the obtained model is equivalent to a theory, from which more concrete models may again be obtained. New opportunities in research can arise from the recognition of the recursive theory-model chain. Via identification of structures, whether free or instantiated, in one's model, opportunities for science arise. An example of this in the context of AC loss modelling can be found in \cite{StenvallHelix}.

\section{Formulating numerical models}

Acknowledging the recursive nature of theories and models does not carry a modeller very far by itself. To utilize models of continuum physics for computer simulations, a \emph{discretization} is required. Naturally, there are many ways to achieve such a discretization and, in addition, different starting points for modelling altogether. In this section, we discuss formulating physics with differential, pointwise quantities and algebraic, macroscopic quantities. Then, we go on to discuss discretizing such a formulation and the theoretical soundess of such a discrete description, i.e. the soundness of \emph{a numerical model} via the \emph{compatibility} of a discretization.

\subsection{Algebraic vs. differential formulation}

\begin{figure}[h]
\center
\includegraphics[scale=0.75]{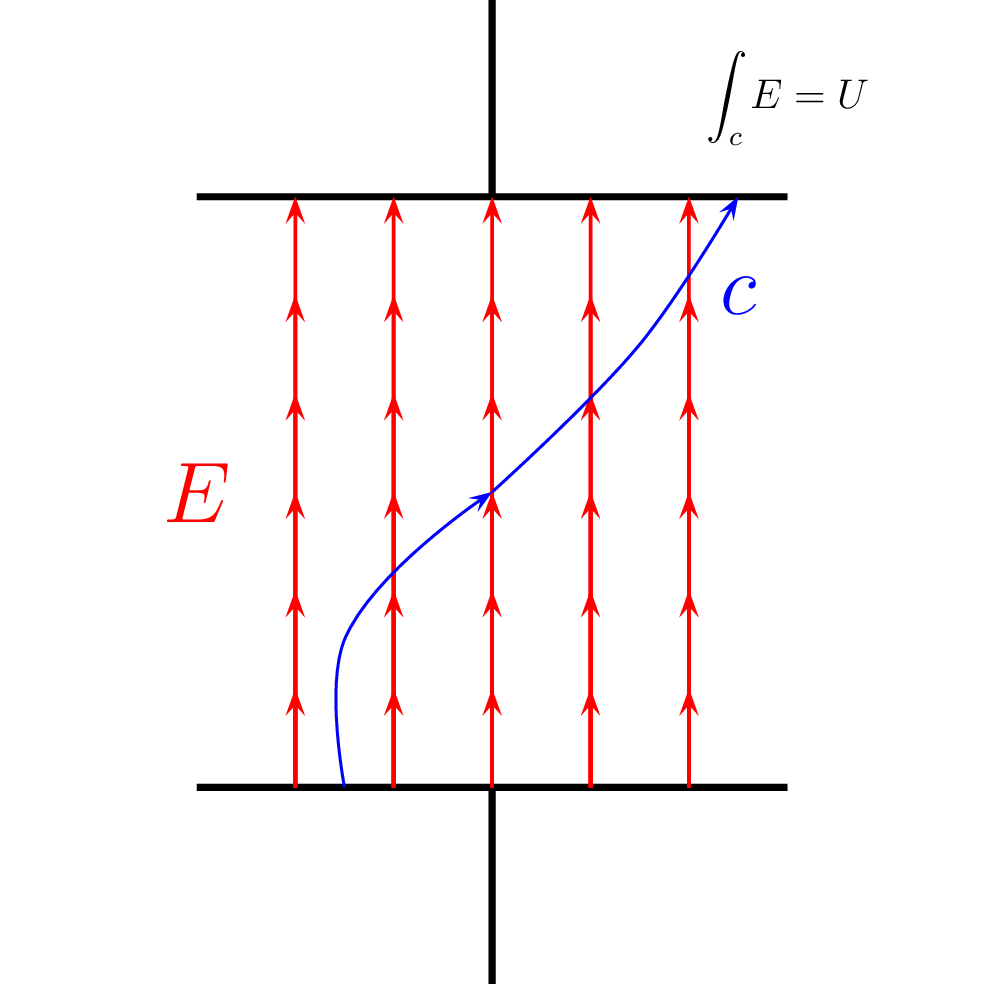}\caption{An integral of the electric field intensity, a differential form or a vector field $E$, over the curve $c$ (1-chain) yields the 1-cochain that represents the voltage $U$ between the end points of $c$.}
\label{fig:Df-Cochain}
\end{figure}

The typical way to formulate physics is to use point-wise quantities, such as vector fields or differential forms, governed by differential equations. This is exactly how we presented Maxwell's theory of electromagnetism in section~2. Another option is to use macroscopic quantities: integrals of vector fields, or macroscopic counterparts of differential forms called cochains \cite{Gross}. See \fref{fig:Df-Cochain} for clarification. For example, Maxwell's theory can be represented either using point-wise quantities and differential equations or macroscopic quantities and algebraic equations. Take, say, the Gauss's law for magnetic field \eqref{Gauss1}: In a similar manner, we may write this as
\begin{equation}\label{GaussCochain}
\Psi(\partial V) = 0 \quad \forall V,
\end{equation}
which states that the magnetic flux $\Psi$ through the boundary of a volume $\partial V$ is zero for all volumes $V$. Here, $\Psi$ is the \emph{magnetic flux 2-cochain} related to the surface (\emph{a 2-chain}) $\partial V$. That is, the 2-cochain $\Psi$ attaches a real number, the flux, to the 2-chain. A $p$-cochain attaches a real number to a $p$-chain: 1-cochains map curves to numbers, 2-cochains map surfaces to numbers and 3-cochains map volumes to numbers. A brief, formal treatment of chains and cochains can be found in the appendix.

These two approaches are, in some sense, inherently different. However, we use them to describe exactly the same phenomena. Moreover, both formalize the idea of continuum: the universal quantifier $\forall$ makes the cochain equations give rise to infinite-dimensional function spaces in a similar manner as differential equations holding point-wise do \cite{Opportunities}.

\subsection{From continuum to discrete: Compatibility of a discrete formulation with the continuum theory}

In the finite element method (FEM), for example, we approach a discrete formulation of a physical problem from the traditional direction by discretizing partial differential equations. That is, we have our model of physics described by such point-wise quantities as vector fields which are governed by partial differential equations. Solving such problems, infinite-dimensional function spaces are encountered, and hence, a finite-dimensionalization i.e. a discretization is needed. Utilizing an algebraic formulation, like e.g. in the \emph{cell method} (CM) \cite{Tonti2014}, one discretizes the infinite-dimensional cochain complex by finding a finite-dimensional subcomplex. As described in \cite{Opportunities}, this is in fact the same thing one does in FEM. Here, we go briefly through these two examples of discretization: FEM on a rather cursory level and CM in more detail. 

Whether we use a differential or an algebraic formulation to begin with, to get reliable predictions the discretization that we make should be theoretically as sound as possible. That is, the discrete description should be compatible with the continuum one in the sense that as much of the structure as possible should be preserved in the discretization. This is our main focus here.

What do we mean by this? Several works explain the idea from different perspectives; see e.g. \cite{CompatibleSpatialDiscretizations} for a collection of such papers. Also, we discussed our views recently on the subject broadly in \cite{Opportunities}. Be that as it may, the main idea boils down to preserving structures: the properties of the differential operators, invariants of the theory, properties of the modelled quantities or properties of the space, for example.

\subsubsection{Compatibility in FEM}

The method we most often utilize to achieve a discrete description of the non-linear magnetoquasistatics problem we are solving to predict the AC losses is FEM. Several formulations have been utilized for this purpose \cite{Brambilla, Hong, StenvallTphi, Morandi-avj, StenvallHPhiPsi, LahtinenCohomology, MikaTphi, T-A}. Reference \cite{LahtinenSUST} presents a comparison of some of the formulations and \cite{acreview} is a comprehensive review of AC loss computation altogether. However, perhaps the most popular formulation is the so-called $H$-formulation, made popular in this context by Brambilla {\it et al.} \cite{Brambilla}.

In (Galerkin) FEM, one first forms the \emph{weighted residual formulation} from the strong form of the partial differential equations by utilizing an inner product in a Hilbert space. Then, the \emph{weak formulation} is obtained through weakening the differentiability requirements via integration by parts. Finally, this weak formulation is discretized by finding a suitable finite-dimensional subspace of the function space from where one is searching the solution for the problem at hand. The choice of the subspace and its basis is of crucial importance for the reliability of the numerical model, as we will discuss below.

We will not delve deeper into the details of FEM within the scope of this paper, as the average reader is likely to be familiar with the method to some extent. For more, see e.g. \cite{Brenner}.

What kind of tools ensure the preservation of structures in a FEM discretization? As an example the $H$-formulation is such a powerful and reliable tool for HTS AC loss modelling, because of its use of \emph{edge elements} or \emph{Whitney 1-forms} \cite{WhitneyForms}. The Whitney form discretization ensures that the continuity properties of the electromagnetic field quantities are preserved in the process. Furthermore, Whitney forms inherit the exterior derivative, which is a generalization of the vector differential operators $\mathrm{grad}$, $\mathrm{curl}$ and $\mathrm{div}$, from the more general setting of differential forms, preserving its properties. Moreover, the so-called \emph{Whitney map} and the \emph{de Rham map}, crucial in the Whitney form discretization, are ways to travel between simplicial cochains (integrals of differential forms on a simplicial mesh) and differential forms (Whitney forms). For formal definitions within a modern treatment see e.g. \cite{Gross}. The properties of these mappings guarantee that the cohomology groups related to differential forms (i.e. in the continuum description) and those related to simplicial cochains (i.e. in the discrete description) are the same for modelling intents and purposes \cite{Opportunities, Gross, Dodziuk76, Muller78}. Thus, important topological properties of the space are preserved in the Whitney form discretization, too.
	
Why care about compatibility? To ensure the \emph{reliability} of the simulations. For instance, the spurious solutions plaguing electromagnetics in the 1980s were disposed of via the introduction of Whitney forms \cite{Bossavit90}. A good example from recent HTS AC loss modelling literature can be given in the context of FEM discretization of the $A$-$v$-$J$-formulation of magnetoquasistatics. In \cite{LahtinenSUST} and \cite{Stenvall-avj} the current density $J$ was discretized using nodal elements, thus rendering $J$ tangentially continuous.\footnote{$J$ has to be spanned with basis functions because of the non-linearity of the power-law visible in the constitutive equation between $E$ and $J$: This way, $\rho$ can be expressed without using the time-derivative of the (unknown) magnetic vector potential $A$, and thus generic time-stepping algorithms may be utilized \cite{Stenvall-avj}.} Having interfaces between materials with differing resistivities, this leads to unphysical current density profiles: The profiles are non-smooth near the material interface \cite{LahtinenSUST}. Antonio Morandi's approach to this problem of incompatibility between the discrete and continuous descriptions was to introduce doubled nodes on the material interface, thus relaxing the condition of tangential continuity \cite{Morandi-avj}. Another option could be to build the numerical model using Whitney 2-forms for $J$, as $J$ is naturally expressed as a differential 2-form in continuum.


Another example of incompatibiltiy due to modelling decisions is how the standard formulation in our field is often utilized: the $H$-formulation. In this formulation, one makes the (physically justified) modelling decision that non-conducting regions are in fact regions with very high resistivity. However, when one sets the Dirichlet boundary conditions for the magnetic field, fixing the magnetic field intensity at the boundary to be the applied external field, one also imposes a net current to flow through the cross-sections of the whole domain. Then, imposing a net current in the conductors separately, a significant current will have to flow in the highly resistive regions of the modelling domain, as discussed e.g. in \cite{LahtinenSUST} and \cite{StenvallHPhiPsi}. Note that this is not as much a matter of incompatibility between the discrete and the continuous as it is a matter of incompatibility between the model and our physical intuition. This can be fixed by taking the self-field into account in the boundary conditions using the Biot-Savart law or with a different discretization altogether, using a truly non-conducting domain and cohomology basis functions to set the net currents \cite{Zermeno, Sotelo, StenvallHPhiPsi, LahtinenCohomology}.

\subsubsection{A case example: Utilizing cohomology in a compatible FEM formulation}

As discussed above, using an $H$-oriented formulation with cohomology basis functions, a compatible FEM formulation is attained. We have discussed the benefits of such a formulation in terms of avoiding leak currents in the air regions in \cite{StenvallHPhiPsi} and presented the formulation in detail and discussed its benefits in terms of faster simulations in \cite{LahtinenCohomology}. Here, we would like to emphasize the ``naturality'' of applying current constraints in an AC loss modelling problem using such a formulation as well as the ramifications on the practicality of full 3D simulations.

\begin{figure}[!ht]
\centering
\includegraphics[scale=0.2]{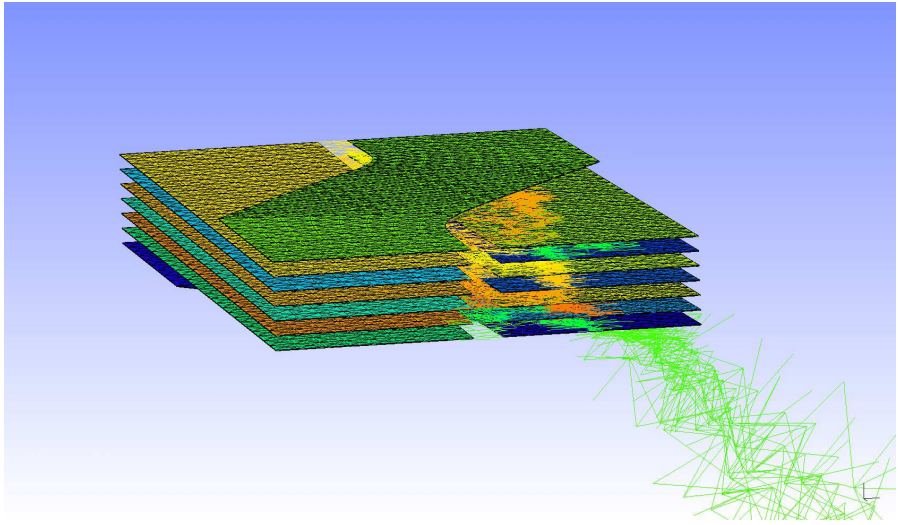}\caption{The edges making up the cohomology basis functions that can be used for setting net currents through the tapes of the Roebel cable mesh. Each of the tapes has a corresponding cohomology basis function reaching from the boundary of the modelling domain up to the boundary of the tape. To minimize the total support of the cohomology basis, the supports of the basis functions overlap. Fixing these cohomology basis functions fixes the integrals of $H$ around the tapes in the $H$-$\varphi$-$\Psi$-formulation.}
\label{fig:roebelCohomology}
\end{figure}

Consider setting current constraints for each of the tapes individually for the Roebel cable mesh utilized in \cite{full3DRoebel}, which represents 1/14th of the transposition length of a Roebel cable consisting of 14 YBCO tapes.\footnote{The mesh is available in Comsol format at the HTS Modelling Workgroup webpage: \url{http://www.htsmodelling.com}} In the $H$-formulation, this is done by introducing an algebraic constraint for each of the currents, thus rendering the equation system to be solved differential-algebraic. As this mesh consists of 708389~edges, posing an AC loss modelling problem in the domain with the $H$-formulation results in a similar number of degrees of freedom, with possible Dirichlet and periodic boundary conditions reducing the number slightly compared to the total number of edges. However, the problem can be set up utilizing the $H$-$\varphi$-$\Psi$-formulation, which makes use of nodal basis functions for scalar potential and cohomology basis in the air regions and uses edge elements only in the conducting regions. Not considering Dirichlet or periodic boundary conditions for simplicity, this results in 175792 total degrees of freedom. There is no need for algebraic constraints as currents are fixed simply by fixing the cohomology basis. Hence, one can solve a pure ordinary differential equation system. A close-up of the cohomology basis for the Roebel cable, which can be used to set the net current constraints for the tapes, is depcited in \fref{fig:roebelCohomology}. The cohomology basis was computed using the cohomology solver implemented within the Gmsh mesh generator software \cite{Pellikka}.

\begin{figure}[!ht]\label{fig:currentsRoebel}
\centering
\subfigure[Current density profile ($||{\bf J}||\times \mathrm{sign}(J_x)$) in the applied field case at the peak of the sinusoidal field ($t = 5$~ms).]{
\includegraphics[scale=0.22]{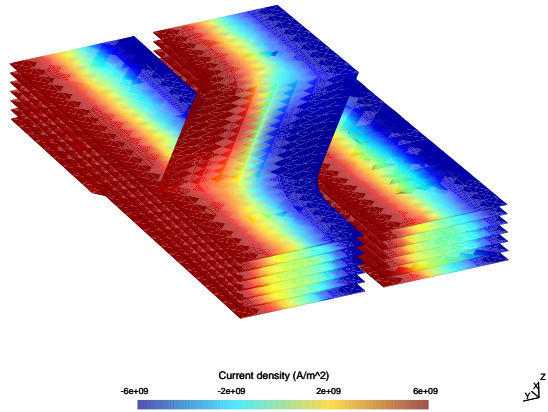}
\label{fig:currentsRoebel1}
}
\subfigure[Current density profile ($||{\bf J}||\times \mathrm{sign}(J_x)$) in the transport current case at the peak of the sinusoidal current ($t = 5$~ms).]{
\includegraphics[scale=0.22]{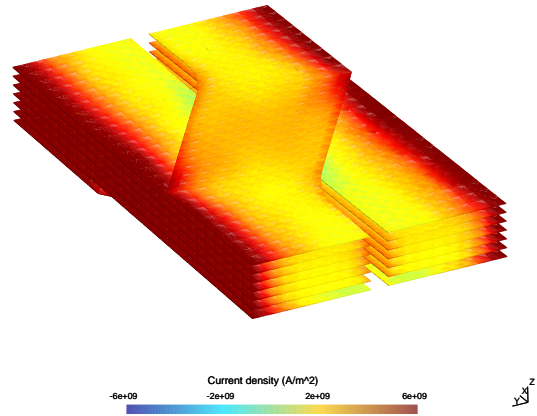}
\label{fig:currentsRoebel2}
}
\caption{Current density profiles.}
\end{figure}

For demonstrative purposes, we solved two cases using this mesh with the $H$-$\varphi$-$\Psi$-formulation utilizing our in-house magnetoquasistatics solver LoST \cite{LahtinenPhD}: an applied field case\footnote{Note that the cohomology basis was not utilized in the applied field case, as we applied no current constraints.} with a sinusoidal magnetic field with frequency $f = 50$~Hz, amplitude $B_\mathrm{app} = 100$~mT and a transport current case with each tape carrying a sinusoidal current of amplitude $I = 0.5 I_\mathrm{c}$, the $I_\mathrm{c}$ for each tape being $100$~A. With cross-sectional dimensions of 10~$\mu$m times 2~mm this yields $J_\mathrm{c} = 0.5 \times 10^{10}$~A/m$^2$. In the power law, we chose $E_\mathrm{c} = 10^{-4}$~V/m and $n = 18$. The current penetration profiles for these example simulations are depicted in \fref{fig:currentsRoebel1} and \fref{fig:currentsRoebel2}.

Hence, utilizing cohomology basis functions has instant ramifications on the practicality of such large 3D simulations via reduced number of degrees of freedom. This is especially the case with such high aspect ratio structures as Roebel cables, as the conducting subdomains remain small compared to the air region required in the modelling. Moreover, setting the current constraints is done naturally by fixing the cohomology basis functions to the desired current values. As there is no need to constrain currents with separate algebraic equations, an ordinary differential equation system instead of a differential-algebraic one may be solved. This makes the spatial discrete formulation compatible with a larger class of time-discretization schemes.

\subsubsection{The cell method: intuitive compatibility with observations}

Since discretizing an algebraic formulation is not a very common methodology utilized in the context of HTS AC loss modelling, we present the main ideas of such a process here using the cell method (CM) as an example. As it does not require the limit process or infinitesimal quantities inherent in the differential equation approaches \cite{Tonti2014}, it has potential for enhancing the understanding of the modelling-related phenomena for students and researchers alike: CM is, in some sense, \emph{compatible with intuition}. Here, we will be brief but we have included an appendix discussing the related background mathematics.

The main idea in CM is to represent the laws of nature using cochains. As an example, Gauss's law for magnetic field in Maxwell's theory of electromagnetism can be expressed as in \eqref{GaussCochain}. Using the magnetic flux differential 2-form $B$, this is the integral form of the Gauss's law, written as
\begin{equation}\label{integralGauss}
\int_{\partial V} B = 0, \quad \forall V,
\end{equation}
where the integral of $B$ can be understood as a representation of a cochain operating on the 2-chain $\partial V$, a boundary of a volume (or a 3-chain) $V$. Or, using vector field formalism, we could write this as 
\begin{equation}\label{integralGaussVector}
\int_{\partial V} \left({\bf B} \cdot {\bf n}\right) d A = 0, \quad \forall V.
\end{equation}
Note however, that \eqref{integralGauss} is more general as the concept of inner product is not required. Similarly, the Amp\`ere's law can be stated using cochains as
\begin{equation}\label{algebraicAmpere}
\Xi(\partial S) = I(S), \quad \forall S,
\end{equation}
stating that the magnetomotive force $\Xi$ around the boundary of a 2-chain $S$ yields the net current $I$ through $S$. Using the magnetic field intensity differential 1-form $H$ and the current density 2-form $J$ this can be written as
\begin{equation}\label{integralAmpere}
\int_{\partial S} H = \int_S J, \quad \forall S.
\end{equation} 

However, it is not necessary to consider cochains as integrals of differential forms or other infinitesimal quantities, and \emph{we can work directly with the macroscopic quantities}: Cochains exist without any reference to differential forms or their existence -- integrals of differential forms are just one representation of cochains. What is important, is that in theories of physics, we want to express the defining properties as \emph{universal} laws: It is necessary that, for example, an equation of the form \eqref{GaussCochain} holds \emph{for all} $3$-chains. But how do we formalize the idea of \emph{all} chains? Obviously, working on a 3-dimensional Riemannian manifold $\Omega$, we must mean that
\begin{equation}\label{algebraicLaw2}
\Psi(\partial V) = 0 \quad \forall V \in C_3(\Omega),
\end{equation}
where $C_3(\Omega)$ is an \emph{infinite-dimensional space of $3$-chains} on the manifold $\Omega$. Combining the spaces of $p$-chains and $p$-cochains with different values of $p$, one obtains \emph{chain and cochain complexes}, as detailed in the appendix. The natural setting for the algebraic formulation of physics, which is the starting point for CM, is thus a chain complex $C_x(\Omega)$ and the corresponding cochain complex $C^x(\Omega)$.

The next step in CM is to construct a mesh -- a cellular mesh complex $M$ and its dual complex $\tilde{M}$ in $\Omega$, from which finite complexes of chains and cochains arise. But what do we mean by a dual complex of a cellular mesh complex and why do we need it? A dual complex of a (primal) chain complex is a complex of chains made of dual cells of the (primal) cells. Intuitively speaking, e.g. in three dimensions the dual of a surface is a line piercing it, and the dual of a volume is a point inside it. Here, the concepts of \emph{inner and outer orientation} as well as \emph{twisted cochains} come into play. Some cochains are associated with an inner orientation of the geometric entity they are related to, while twisted cochains are associated to outer oriented chains. As an example, the electric field is naturally described on a path with inner orientation, a sense of positive propagation direction, while electric current flows through a surface, and hence, its outer orientation, or positive crossing direction, is meaningful. Thus, electric currents are twisted cochains. An inner oriented curve piercing an orientable surface defines a positive crossing direction for the surface: There is an intimate connection between inner and outer orientations of cells of different dimensions. For formal definitions, see appendix.

\begin{figure}[!ht]
\centering

\begin{pspicture}(0,0)(12,8)


\psline(0,0)(8,8)
\psline(0,4)(4,8)
\psline(4,0)(12,8)
\psline(8,0)(12,4)

\psline(0,4)(4,0)
\psline(0,8)(8,0)
\psline[arrowsize=6pt]{->}(4,8)(12,0)
\psline(8,8)(12,4)

\psdot[dotsize=3pt](6,4)
\psdot[dotsize=3pt](8,2)
\psdot[dotsize=3pt](8,6)
\psdot[dotsize=3pt](4,6)
\psdot[dotsize=3pt](2,4)
\psdot[dotsize=3pt](4,2)
\psdot[dotsize=3pt](10,4)
\psdot[dotsize=3pt](0,2)
\psdot[dotsize=3pt](0,6)
\psdot[dotsize=3pt](2,8)
\psdot[dotsize=3pt](6,8)
\psdot[dotsize=3pt](10,8)
\psdot[dotsize=3pt](2,0)
\psdot[dotsize=3pt](6,0)
\psdot[dotsize=3pt](10,0)
\psdot[dotsize=3pt](12,2)
\psdot[dotsize=3pt](12,6)

\psline[linestyle=dashed](0,6)(2,8)
\psline[linestyle=dashed](0,2)(6,8)
\psline[linestyle=dashed](2,0)(10,8)
\psline[linestyle=dashed](6,0)(12,6)
\psline[linestyle=dashed, linecolor=red, linewidth=2.0pt](10,0)(12,2)

\psline[linestyle=dashed](0,2)(2,0)
\psline[linestyle=dashed](0,6)(6,0)
\psline[linestyle=dashed](2,8)(10,0)
\psline[linestyle=dashed](6,8)(12,2)
\psline[linestyle=dashed](10,8)(12,6)

\end{pspicture}

\caption{\label{fig:primalDualMesh}A partial depiction of a cellular mesh complex and its dual, suitable for cell method, on a 2-dimensional manifold $\Omega$. Solid and dashed lines represent 1-cells of the primal and dual mesh, respectively. The 0-cells of the dual mesh have been identified, as well. The inner orientation of one primal 1-cell, i.e. propagation direction, in the lower right corner is indicated. Note how it yields a sense of positive crossing direction (outer orientation) for the dual 1-cell it pierces, indicated by red color and thickening, making it meaningful to e.g. compute fluxes across the dual 1-cell.}
\end{figure}
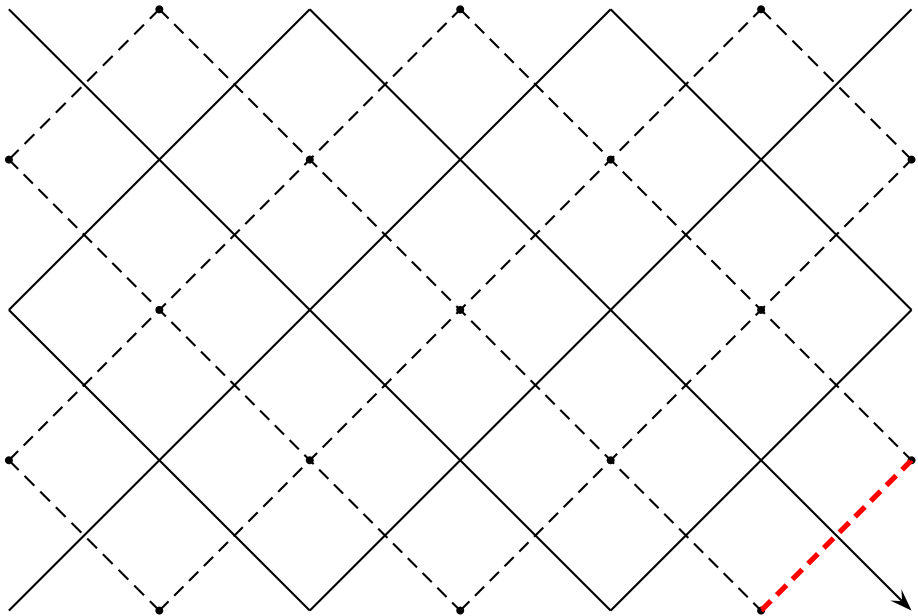

Hence, one associates an outer oriented $(n-p)$-cell of the dual complex $\tilde{M}$ to each inner oriented $p$-cell of the primal complex $M$ on an $n$-dimensional $\Omega$. The inner orientation of the primal complex induces the outer orientation of the dual complex. This primal-dual pairing of mesh complexes gives us also a way to connect $p$-cochains with twisted ones and talk about e.g. constitutive equations of physical theories, utilizing the Riemannian structure on $\Omega$ \cite{Tonti2014}. So we have the cellular mesh complex, which gives rise to a finite chain complex $C_x(M, \Omega)$ and a finite cochain complex $C^x(M, \Omega)$. Moreover, the outer oriented dual complex gives rise to a finite outer oriented chain complex $\tilde{C}_x(\tilde{M}, \Omega)$ and a finite twisted cochain complex $\tilde{C}^x(\tilde{M}, \Omega)$. These complexes are then used for formulating physics in a finite way. These ideas are summarized and clarified in Table~\ref{table:primalDual} and \fref{fig:primalDualMesh}. For more details, see e.g.  \cite{Tonti2014, Keranen2011, Kreeft}.

\begin{table}
\centering
\begin{tabular}{cc}
Primal	 &	Dual \\ \hline
$M$			 &	$\tilde{M}$ \\ 
$C_x(M, \Omega)$ & $\tilde{C}_x(\tilde{M}, \Omega)$ \\
$C^x(M, \Omega)$ & $\tilde{C}^x(\tilde{M}, \Omega)$ \\
$p$-cell				 & $(n-p)$-cell	\\
Inner orientation & Outer orientation \\
\hline
\end{tabular}
\caption{\label{table:primalDual}Correspondence between primal and dual cell complexes and arising (co)chain complexes.}
\end{table}

This is the \emph{meshing process} present in CM. We do not require the equations to hold \emph{for all} possible $p$-chains anymore, but only for those arising from the primal-dual cellular mesh complex pair. As the cellular mesh complexes, and thus also the resulting chain and cochain complexes are finite, CM arrives at a finite number of equations. As an example, in magnetoquasistatics, we would require that the electric currents through the 2-chains of the dual mesh equal the magnetomotive forces around their boundaries. This is the Amp\`ere's law. Or Gauss's law \eqref{GaussCochain} would now state that the magnetic flux cochain is zero for all outer oriented bounding 2-chains in the dual mesh. This process thus renders \eqref{GaussCochain} and \eqref{algebraicAmpere} on 3-dimensional $\Omega$ to 
\begin{equation}\label{algebraicGaussCM}
\Phi(\partial V) = 0, \quad \forall V \in C_3(M,\Omega),
\end{equation}
and
\begin{equation}\label{algebraicAmpereCM}
\Xi(\partial S) = I(S), \quad \forall S \in \tilde{C}_2(\tilde{M},\Omega).
\end{equation}
Similarly, we would present any equation in terms of these finite (co)chain complexes arising from the cellular mesh complexes and arrive at a discrete formulation.

In time-dependent problems, such as those of magnetoquasistatics encountered in AC loss modelling, naturally also a cell complex in time is required. This can be achieved utilizing a leap-frog type scheme. For the detailed formulation of CM for magnetoquasistatics, see \cite{Bellina}. To the authors' knowledge, this approach has not been tried in superconductor modelling. Hence, the possible computational benefits in our field are yet unknown. This could also be worth a try regarding the pedagogical aspects of such an algebraic formulation \cite{Tonti2014}.

Indeed, CM is a discrete description of continuum physics obtained from an algebraic formulation of physics. First of all, it may be argued that that such a formulation is more directly compatible with our intuition about measurements: We can not measure the differential quantities but we must measure macroscopic ones. Moreover, the orientations and the geometric nature of physical quantities become more transparent. CM is intuitive and compatible with observations in this sense \cite{Tonti2014}.

However, the preassumed continuum space is still present in the context of CM, too. Thus, the question remains: in which sense is this description of physics compatible with the continuum description it mimics? As obtaining results in arbitrary spatial points requires interpolation, CM is not free of the compatibility and convergence issues exhibited by FEM. Moreover, it lacks a similar canonical description of metric aspects that are manifest in the constitutive laws, as opposed to the Whitney form discretization combined with the Galerkin-Hodge process in FEM, as discussed in \cite{Opportunities}.

\subsection{Concluding remarks}

Whether one uses a differential or an algebraic formulation of continuum physics, it is crucial for the reliability of simulations to have a numerical model that preserves the structures of the continuum model it mimics as well as possible. We call this \emph{compatibility} of the discrete description with the corresponding continuum description. This issue is currently underemphasized in our field. Furthermore, an algebraic formulation of physics is in some sense more compatible with intuition and observations. Starting from an algebraic formulation, numerical models like CM can be obtained. Such a formulation has pedagogical value and can be considered as an alternative to differential equations based approach. Currently, we do not know if there are computational benefits to such formulation in non-linear magnetoquasistatics but this should be studied. 

\section{Model validation: relating simulations and experiments}

The compatibility of a discrete description with the continuum model is crucial for the \emph{internal} consistency of the numerical model. Moreover, algebraic formulations of continuum physics provide an intuitive and measurement-like interpretation for us. However, none of this says anything about compatibility with measurements. This is when we begin to discuss \emph{model validation}. 

Modelling need not even take a stand if a model is \emph{right} or \emph{wrong}. Instead, it is crucial to study the \emph{usefulness} of models. If we, as modellers, predict something that deviates from what we observe, it does not mean that the physical theory we are using is \emph{wrong}; indeed, another model of the same theory may yield predictions more compatible with the observations. Moreover, we have built the physical theory inside a mathematical one, and obviously, a mathematical theory is consistent without any reference to physical reality. It is not significant to ask if a model or a theory is right or wrong, as long as it is self-consistent. 

Hence, to validate a model cannot mean to test its rightfulness. But if it is not that, then what is it exactly? In \cite{Oberkampf}, the authors differentiate between \emph{verification} and \emph{validation}. They define verification as the assesment of the accuracy of the solution (this is related to our term \emph{compatibility} in section~3) and validation as the assesment of the accuracy of a simulation by comparison with experiments. Hence, their term verification is related to benchmarking with e.g. analytical solutions and validation to benchmarking with measurement data. 

Let us approach model validation in the context of HTS AC loss modelling via a literature example.

\subsection{Literature example: Ripple field losses in direct current (DC) biased superconductors}

Ripple field losses in HTS biased with direct current (DC) have been investigated in some recent works \cite{LahtinenRipple}, \cite{Lahtinen2013}, \cite{XuGrilli}. In terms of AC loss modelling, such situations exhibit very long transients compared to typical AC situations, such as losses due to AC current, AC field or combinations of those. Additionally, it seems that different types of behaviour can be predicted depending on the chosen $E$-$J$-relation in the model \cite{LahtinenRipple}. The fact that AC ripple fields in applications such as motors and generators tend to be small compared to the DC bias accentuates the difference between e.g. the power law (PL) and the critical state model (CSM): PL predicts a non-zero loss for any value of current density while CSM only predicts losses above $J_c$. This leads to spatial homogenization of the current density profiles over time in PL based simulations, further complicating the long transient behaviour, as discussed in \cite{LahtinenRipple}.

In \cite{LahtinenRipple}, two models of the magnetoquasistatic Maxwell's theory, the PL model and CSM were utilized to predict losses in a DC biased ReBCO tape in AC ripple fields. Furthermore, the predictions of the models were compared against experimental data. Or, as one often says, the \emph{models were validated} against measurements. What could be inferred? The authors report that the qualitative and quantitative agreement between the predictions of PL and CSM was good in a wide range. Furthermore, qualitative and quantitative agreement with measurements was also good in a wide range of situations. However, discrepancies at low AC fields with significant DC bias were observed, prominently when CSM was utilized. On the other hand, the PL based model exhibited discrepancies especially in terms of magnetization loss. 

A lot of uncertainty is of course related to such simulations and measurements. The authors note that there was a possible degradation of the critical current density $J_\mathrm{c}$ near the tape edges that was not taken into account in the simulations. Also $J_\mathrm{c}$ was possibly under-estimated at low fields, and the potential current sharing with the stabilization layer was not taken into account. Furthemore, the noise-to-signal ratio at low fields is not good, adding further uncertainty to the results. Finally, the discretization methods were in fact of totally different nature, and the mutual compatibility of these discrete descriptions was not detailedly analyzed.

\subsection{Concluding remarks}

As reported in \cite{LahtinenRipple}, one model can reflect measurements better in a range of situations while one in another. So were the models validated in \cite{LahtinenRipple}? At least not in the sense that one model or another would have been proven right or wrong. In AC loss modelling, we tend to look at the double integral of the loss density $P = {\bf E} \cdot {\bf J}$
\begin{equation}\label{ACLoss}
\int_T \int_{\Omega_{\mathrm{sc}}} P \rmd V \rmd t
\end{equation}
over the superconducting regions $\Omega_{\mathrm{sc}}$ and cycle $T$ of the AC quantity and compare models and measurements solely based on this real number. Such lumping can easily hide some essential properties of the utilized models. A theory-model system is or is not internally consistent regardless of model validation. Moreover, model validation like this does not validate a model in terms of compatibility of the discretization, which is rarely discussed in the context of model validation. To conclude, model validation demonstrates the applicability of the modelling methodology in some particular cases via comparison with measured data.

\section{Summary and conclusions}

Methodological and terminological issues of HTS AC loss modelling have been addressed in recent workshops and conferences. Discussions have indicated a need for addressing such questions in more formal format. In this paper, we have started this discussion and considered some of these issues with (subjectively) of high importance. We discussed the general recursive nature of models and theories and the formulation of numerical models, emphasizing the \emph{compatibility} of such models with the continuum theory as well as our intuition. Finally, we discussed the relation of modelling and experiments, scrutinizing the term \emph{model validation} via an example from the literature.

\emph{Validation} of a model via comparison with experiments is of crucial importance in the endeavour to produce predictions of AC losses in HTS. However, we feel that the \emph{verification} of the reliability of the results in terms of theoretical soundness is currently underpresented in our field and should be addressed more carefully. Approaching numerical modelling also from an algebraic point of view, in addition to the traditional differential equations approach, could broaden the horizons of researchers working on AC loss modelling via increased intuition and the necessarily different mathematical machinery, which is after all inherently present in the discretized versions of differential equation based formulations, too.

To conclude and summarize the ideas presented in this work, we would like to end this paper with four conjectures regarding modelling in general, but learned in the context of and especially valid for HTS AC loss modelling.

\begin{myconjecture}

Theories and models should be considered as a recursive system; Whether a mathematical description of nature should be considered a model or a theory depends on the abstraction level from where one is looking.

\end{myconjecture}

\begin{myconjecture}

While modelling, it is important to realize what there is to instantiate and what is one (or more) abstraction level(s) up.

\end{myconjecture}

\begin{myconjecture}

The compatibility of a discrete formulation with the continuum formulation is of great importance in achieving reliable predictions: the properties of the continuum description should be preserved in a discretization as well as possible. 

\end{myconjecture}

\begin{myconjecture}

``Model validation'' as it is understood, never fully validates a model. Demonstrating the applicability of a modelling methodology via comparison with measurements is necessary, but there are further aspects to the validity of a model that should be considered.

\end{myconjecture}

\noindent We hope these conjectures serve as a springboard for further discussions within the numerical superconductor modelling community in general and among HTS AC loss modellers in particular.

\section*{Acknowledgment}

This research was partially supported by The Academy of Finland project [287027]. The authors would like to thank the organizers and attendees of the Sixth International Workshop on Numerical Modelling of HTS held in Caparica, Portugal, 26-29 June 2018, for providing a fertile soil for this paper to grow from. The authors also extend their gratitude to Francesco Grilli and Victor Zermeno for providing the Roebel cable mesh used in the simulations of this paper.

\appendix
\setcounter{footnote}{0}
\section{Brief introduction to chains and cochains}

This is a brief introduction to chains and cochains. For further details, we refer the reader to e.g. \cite{Whitney}. This appendix is mainly based on the thesis by Ker\"anen \cite{Keranen2011}.

\subsection{Cells}

Often, a numerical solution method relies on computations on a \emph{mesh} or \emph{grid} consisting of simple geometrical entities. A \emph{cell} is a model for these simple entities. We shall thus work within this cellular context on a Riemannian manifold $\Omega$.

Recall that given a finite set $X$ of points in the coordinate space $\mathbb{R}^n$, a {\bf convex cell} is its convex closure.\footnote{For $X \subset \mathbb{R}^n$, its convex closure is a closed minimal convex set containining $X$.} Its dimension is defined as the dimension of the smallest affine subspace that contains $X$. Then, a $p$-dimensional {\bf polyhedral cell} or a {\bf $p$-cell} for short, is an equivalence class $c$ of pairs $(\chi, C)$, where $\chi: \Omega \rightarrow \mathbb{R}^n$ is a chart on $\Omega$ and $C$ is a convex cell in the range of $\chi$. In the classification, we deem pairs $(\chi_1, C_1)$ and $(\chi_2, C_2)$ equivalent if and only if $\chi_2 \circ \chi_1^{-1}(C_1) = C_2$. The dimension $p$ of a $p$-cell is that of its convex cells.

A polyhedral cell inherits its orientation or {\bf inner orientation} from its convex cells, which in turn inherit their orientation from the affine space where they reside in. A polyhedral cell is an {\bf inner oriented cell}. The concept of outer orientation, however, is also important in physics. An {\bf outer oriented cell} is defined as an equivalence class of pairs of a polyhedral $p$-cell $c$ and global orientation $O$ on the manifold $\Omega$.\footnote{A global orientation of $\Omega$ assigns an orientation to each tangent space $\TxM$ of $\Omega$.} In the classification, ($c_1, O_1$) and ($c_2, O_2$) are equivalent if and only if the $p$-cells are equal as sets and the inner orientations of $c_1$ and $c_2$ match precisely when $O_1 = O_2$. Loosely speaking, the inner orientiation of a cell is a private property of the cell and its outer orientation is a more public one, as it depends on the orientation of the manifold it lives on.

\subsection{Chains and cochains}
 
Next, we want to construct a linear space of collections of $p$-cells, to talk about, e.g., general paths and surfaces and basic algebraic operations for them. Moreover, to such collections, we want to associate real numbers, and construct a linear space of such associations, to be able to model physical quantities.

A {\bf (polyhedral) chain} of dimension $p$ or a {\bf $p$-chain} is an equivalence class of formal sums\footnote{Such objects are called \emph{formal} sums, when one writes the combinations of objects forming them as ``sums'' using the plus sign, even though a sum operation might not actually be defined for the ``summands''.} 
\begin{equation}\label{formalSum}
\sum_{} \alpha(c_i)c_i
\end{equation}
of $p$-cells $c_i$, where $\alpha$ maps $p$-cells to real numbers so that the complement of its kernel is finite. That is, each chain can be represented with a finite amount of information. In the equivalence classification, two such formal sums are equivalent if and only if their difference belongs to the \emph{zero class of formal sums}.\footnote{The zero class contains those formal sums that are intuitively zero, such as the zero formal sums and the formal sums of cells that are equal as sets but their orientations are opposite. For more, see e.g. \cite{Keranen2011}.} This way, we ensure that our intuition of the sameness of geometrical objects that the chains model is preserved, as many formal sums of cells can represent the same geometrical object. The quotient space thus constructed is the {\bf vector space of $p$-chains on $\Omega$}, $C_p(\Omega)$. In an analogous manner, we can define also the {\bf vector space of outer oriented $p$-chains on $\Omega$}, $\tilde{C}_p(\Omega)$. Note that these spaces are infinite-dimensional vector spaces since the set of all polyhedral cells on a manifold is infinite.

It is often necessary to talk about a boundary of a geometrical object. To define the boundary of a chain, we first need a definition for the boundary of a cell. The boundary operator $\partial$ for cells maps $p$-cells to $(p-1)$-chains. As an intuitive example, one can think of a cube (represented by a cell) in $\mathbb{R}^3$, the boundary of which consists of a combination of six squares (represented by a chain). As a convex cell $C$ is the convex closure of a finite set $X$, {\bf the boundary of a convex cell} is the set $\partial C$ consisting of those points that lie in the intersection of the closure of $X$ and the complement of the interior of $X$. {\bf The boundary of a $p$-cell} can then be defined through its representative. The $(p-1)$-cells forming the boundary of a $p$-cell are called {\bf faces} of the $p$-cell. Dually, those $(p+1)$-cells that have the $p$-cell as a common face are called {\bf cofaces} of the $p$-cell. The {\bf boundary operator for $p$-chains} is linearly extended from the boundary operator for cells. It is a mapping $\partial: C_p(\Omega) \rightarrow C_{p-1}(\Omega)$, such that
\begin{equation}\label{boundaryChain}
\partial (\sum \alpha_i c_i) = \sum \alpha_i \partial c_i,
\end{equation}
taking $p$-dimensional chains to their $(p-1)$-dimensional boundaries. As intuition suggests, $\partial \circ \partial = 0$. Finally, A {\bf dual cell} of a $p$-cell is such that its cofaces are the dual cells of the boundary of the $p$-cell.

{\bf Cochains} of dimension $p$ on the manifold $\Omega$, or {\bf $p$-cochains} for short, form the dual space of $C_p(\Omega)$, denoted by $C^p(\Omega)$. That is, they are continuous linear mappings from $C_p(\Omega)$ to $\mathbb{R}$. Moreover, through equivalence classification of pairs of $p$-cochains and global manifold orientations, we can define {\bf twisted $p$-cochains}, which take outer oriented $p$-chains to real numbers.\footnote{For example, heat flux through a surface is a twisted 2-cochain: It takes an outer oriented 2-chain (a surface with an assigned positive crossing direction) to a real number.} The vector space they form is denoted by $\tilde{C}^p(\Omega)$.

For chains we have the boundary operator $\partial$, and dually, for cochains we define the {\bf coboundary operator} as a linear mapping $\rmd: C^p(\Omega) \rightarrow C^{p+1}(\Omega)$. Its operation on a $p$-cochain $\Psi$ is defined via
\begin{equation}\label{coboundary}
\rmd \Psi(\Gamma) := \Psi(\partial \Gamma),
\end{equation}
where $\Gamma$ is a $(p+1)$-chain. Also for $\rmd$ it holds that $\rmd \circ \rmd = 0$. As a familiar example, via integration, the exterior derivative $\rmd$ of differential forms is an instance of the coboundary operator, as the notation also suggests.\footnote{The exterior derivative $\rmd$ takes differential $p$-forms to $(p+1)$-forms on a manifold $\Omega$. Differential $p$-forms are, by definition, integrated over $p$-chains, and integrating a $p$-form over a $p$-chain yields a real number. Now Stokes' theorem states that $\int_\Gamma \rmd \eta = \int_{\partial \Gamma} \eta, \quad \forall \Gamma$, where $\Gamma$ is a $(p+1)$-chain and $\eta$ is a $p$-form. Thus, $\rmd$ takes the $p$-cochain represented by the right-hand side of the equation to the $(p+1)$-cochain on the left-hand side.}

Finally, to model quantities related to geometric objects of different dimensions, we combine the (co)chains of different dimension in a single structure using the (co)boundary operator, a {\bf (co)chain complex}. These are expressed as
\begin{equation}\label{chainComplex}
0 \rightarrow C_n(\Omega) \stackrel{\partial}{\rightarrow} C_{n-1}(\Omega) \stackrel{\partial}{\rightarrow}... \stackrel{\partial}{\rightarrow} C_0(\Omega) \rightarrow 0
\end{equation}
and
\begin{equation}\label{cochainComplex}
0 \leftarrow C^n(\Omega) \stackrel{\rmd}{\leftarrow} C^{n-1}(\Omega) \stackrel{\rmd}{\leftarrow}... \stackrel{\rmd}{\leftarrow} C^0(\Omega) \leftarrow 0.
\end{equation}
Similarly, one constructs complexes from outer oriented chains and twisted cochains.

\end{document}